\documentclass[twocolumn,american]{revtex4-2}
\usepackage[T1]{fontenc}
\usepackage[latin9]{inputenc}
\usepackage{xcolor}
\usepackage{babel}
\usepackage{booktabs}
\usepackage{amsmath}
\usepackage{graphicx}
\PassOptionsToPackage{normalem}{ulem}
\usepackage{ulem}
\usepackage[pdfusetitle,
 bookmarks=true,bookmarksnumbered=false,bookmarksopen=false,
 breaklinks=false,pdfborder={0 0 1},backref=false,colorlinks=false]
 {hyperref}

\makeatletter

\newcommand*\LyXZeroWidthSpace{\hspace{0pt}}
\providecommand{\tabularnewline}{\\}
\providecolor{lyxadded}{rgb}{0,0,1}
\providecolor{lyxdeleted}{rgb}{1,0,0}
\DeclareRobustCommand{\mklyxadded}[1]{\bgroup\color{lyxadded}{}#1\egroup}
\DeclareRobustCommand{\mklyxdeleted}[1]{\bgroup\color{lyxdeleted}\mklyxsout{#1}\egroup}
\DeclareRobustCommand{\mklyxsout}[1]{\ifx\\#1\else\sout{#1}\fi}
\DeclareRobustCommand{\lyxadded}[4][]{\texorpdfstring{\mklyxadded{#4}}{#4}}

\makeatother

\begin{document}
\title{Shell Models on Recurrent Sequences: Fibonacci, Padovan and Other
Series}
\author{L. Manfredini, {\"O}. D. G\"urcan}
\affiliation{Laboratoire de Physique des Plasmas, CNRS, Ecole Polytechnique, Sorbonne
Universit\'e, Universit\'e Paris-Saclay, Observatoire de Paris,
F-91120 Palaiseau, France}
\begin{abstract}
A new class of shell models is proposed, where the shell variables
are defined on a recurrent sequence of integer wave-numbers such as
the Fibonacci or the Padovan series, or their variations including
a sequence made of square roots of Fibonacci numbers rounded to the
nearest integer. Considering the simplest model, which involves only
local interactions, the interaction coefficients can be generalized
in such a way that the inviscid invariants, such as energy and helicity,
can be conserved even though there is no exact self-similarity. It
is shown that these models basically have identical features with
standard shell models, and produce the same power law spectra, similar
spectral fluxes and analogous deviation from self-similar scaling
of the structure functions implying comparable levels of turbulent
intermittency. Such a formulation potentially opens up the possibility
of using shell models, or their generalizations along with discretized
regular grids, such as those found in direct numerical simulations,
either as diagnostic tools, or subgrid models. It also allows to develop
models where the wave-number shells can be interpreted as sparsely
decimated sets of wave-numbers over an initially regular grid. In
addition to conventional shell models with local interactions that
result in forward cascade, a particular helical shell model with long
range interactions is considered on a similarly recurrent sequence
of wave numbers, corresponding to the Fibonacci series, and found
to result in the usual inverse cascade.
\end{abstract}
\maketitle

\section{Introduction}

Turbulence represents one of the most challenging problems in nonlinear
physics, with applications ranging from fluid dynamics and plasma
physics to ocean currents and atmospheric dynamics. One of its key
intrigues is that despite the presence of spatio-temporal chaos across
a wide range of scales it nonetheless exhibits order, specifically
in the form of self-similarity of its hierarchy of spatial structures,
encapsulating the concept of restored statistical symmetry \citep{frisch,falkovich:09}.
Consequently, one of the main lines of inquiry in turbulence research,
is that of the study of the turbulent fluid as a non-equilibrium steady
state with energy injection and dissipation at well separated scales
and a nonlinear transfer in between those, in a so-called inertial
range, resulting in power law solutions for the turbulent spectra
and intermittency of the statistics of its fluctuations \citep{rose:78,frisch}.

One of the main branches of academic research on turbulence is dedicated
to development of reduced models \citep{bohr_Jensen_Paladin_Vulpiani_1998},
focused on various reduction strategies hoping to find one that reduces
the system to a tractable problem without loosing any essential features.
In this context, shell models \citep{desnianskii:74,yamada:88,lvov:98,biferale:03,ditlevsen:2010}
appear as one of the simplest yet most prominent examples of such
models. Notably, shell models have been used for understanding certain
complex aspects of turbulence, such as multifractality \citep{biferale1999,benzi:93,biferale:97,ray:08},
which are too difficult to tackle using the full Navier-Stokes equations.
Additionally, these low-dimensional dynamical systems have shown their
versatility in modeling other systems like rotating turbulence \citep{hattori:04},
passive scalar advection \citep{jensen1992} and convection \citep{kumar:15}.
Beyond fluid turbulence, shell models has been used to study the nature
of cascades in magnetohydrodynamic (MHD) turbulence \citep{plunian:07,plunian:2013},
including the transition from weak to strong cascade in reduced MHD
\citep{verdini:12}. In the context of fusion plasma, they have also
been applied to study the L-H transition using multi-shell models
\citep{berionni:17}. 

In their general form, shell models consist of a set of ordinary differential
equations for a set of variables corresponding to a group of wave-numbers
that retain only the essential characteristics of the original equations---namely,
the quadratic nonlinearity, the scale invariance, and a resulting
set of quadratic conserved quantities. These models can reproduce
some of the key features of fully developed 3D turbulence, such as
the power law exponents of the turbulent cascades and the intermittency
of their fluctuations \citep{kadanoff1995,pisarenko:93}. They are
usually written in a desired form where the coefficients are computed
using the constraints imposed by the conservation laws instead of
being derived from the original system of equations.

Alternatively, shell models can be seen as a simultaneous reduction
of the wave-number domain using a set of logarithmically spaced wave-numbers
$k_{n}\sim k_{0}g^{n}$ , and the nonlinear interaction operator by
reducing the number of possible interactions in the convolution integral
to a finite set of -usually local- interactions. Historically, a complex
variable $u_{n}$ is used to represent the Fourier modes in the shell
$n$ which contains the wave-numbers between $k_{n}$ and $k_{n+1}$
with a constant inter-shell ratio $g$, -usually chosen to be equal
to 2, which would interact non-linearly with the two neighboring shells
on both sides. However choosing $g=2$ is somewhat problematic, because
since it is impossible to have $\mathbf{k}_{n}+\mathbf{k}_{n+1}+\mathbf{k}_{n+2}=0$
with $g=2$, which means that the cell centers do not interact, and
the interactions that are described in such a shell model come only
from the parts of the shells that are close to the boundaries between
the consecutive shells. In other words $g=2$ produces shells that
are in fact too large to keep track of where the energy actually goes.

This is also not ideal for establishing the connection between the
shell model and Fourier space decimation as in the case of spiral
chain models for example \citep{gurcan:19}, which incidentally gives
exactly the same equations as shell models, but only with $g$ values
that permit triadic interactions, such as $g=\sqrt{\varphi}$ where
$\varphi$ is the golden ratio. Note that g=$\varphi$ is the largest
possible value for shell spacing if one wants to use Fourier space
decimation with actual triadic interaction between cell elements.

More generally, the use of logarithmic discretization, used in shell
models, but also in other reduced descriptions such as logarithmically
discretized models \citep{gurcan:16a}, log-lattice models \citep{campolina:2018,campolina:2021,costa:2023}
or spiral chains \citep{gurcan:19}, allows covering a large range
of scales using a small number of degrees of freedom, providing an
important advantage over direct numerical simulations on regular Fourier
space grids, such as those in pseudo-spectral codes. However apart
from $g=2$, which we argue to be dubious, the discretized wave-numbers
$k_{n}$, do not fall on regular grid elements. This means that if
we start from a regular Fourier space grid, and construct a reduced
model of this type, which has the mathematical form of a shell model
(including log-lattice models), the shell elements which would have
$k_{i}=\left\{ 1,\varphi,\varphi^{2},\varphi^{3},\cdots\right\} $
etc. do not fall on regular grid points. In order to remedy this,
here we propose a shell model like discretization, but on a set of
wave-numbers that follow a recurrent sequence such as the Fibonacci
series (skipping the first two elements in order to have the same
structure as a shell model): $k=\left\{ 1,2,3,5,\cdots\right\} $.
Note that eventually the Fibonacci sequence converges to the above
geometric sequence, and we get the usual progression with spacing
approximately equal to the golden ratio $g\approx\varphi$ for large
values of $n$.

Even though the Fibonacci series allows each wave-vector in a sequence
to be the sum of the two precedent wave-vectors, and as such it allows
each shell to interact with every part of the two consecutive shells
as opposed to the $g=2$ case, it does not \emph{really} allow triadic
interactions between cell centers since the area of such a triad and
thus its interaction coefficient vanishes. In order to remedy this,
we have also considered the Padovan sequence. Three consecutive wave-numbers
that follow a Padovan sequence can actually form triads, (asymptotically
with the angles, $\theta_{kp}\approx97.04^{{^\circ}}$, $\theta_{kq}\approx34.44^{{^\circ}}$
and $\theta_{pq}\approx48.52^{{^\circ}}$), even though a chain of
such triads does not \emph{a priori} fall on regular grid points,
since it is the absolute values of the wave-numbers that are integers.
Note that if one generalizes this idea to two or three dimensions
for example using log-lattice models or spiral chains, the wave-number
sequences would represent separate components $k_{n}^{\left(x,y,z\right)}$
(or $k_{n}^{\left(r,\theta\right)}$ etc.). For example, in the case
of log-lattices, such wave-number sequences would naturally form triads
since it is the components that satisfy the recurrence relations.
In this sense, even though we focus on shell models here, the idea
of using recurrent sequences to replace logarithmic scaling is very
general and applicable to all kinds of models and systems, including
log-lattices \citep{campolina:2021} where the components can be written
as $g^{n}$ where $g$ is a special ratio such as the golden ratio,
or the plastic ratio (by the way, one can add the supergolden ratio
to that list) or the nested polyhedra models \citep{gurcan:17,gurcan:18}
where the $x,y,z$ components of the vertex positions of the initial
regular polyhedra can be rearranged to be written in terms of various
combinations of $0,1,$$\pm\varphi$, $\pm\varphi^{2}$ so that exact
triadic interactions are possible. In all these cases, considering
a recurrent sequence will transform the logarithmic scaling to the
corresponding recurrent series, while keeping the triadic interactions
intact.

We observe that a scaling of $g=\sqrt{\varphi}$ also allows to form
actual triangles, with a non-zero area, leading us to also consider
the sequence of numbers that consist of the square roots of the Fibonacci
numbers rounded to the nearest integer as an alternative sequence.
In contrast to those series that satisfy additive recurrence relations,
this series satisfy a recurrence relation for the squares of the wave-number
magnitudes. As such this series would be more suitable for a logarithmically
discretized model (where the discretization is logarithmic in $k$
and linear in $\theta$ in polar coordinates) as opposed to a log-lattice
where the discretization is logarithmic in Cartesian coordinates. 

We further considered variations and combinations of these sequences,
and generally observed that shell models on recurrent integer sequences
behave virtually identically to conventional shell models. Some of
the proposed sequences may be linked to particular elements of a regular
grid based on different geometric constructions. For example, the
Fibonacci spiral can be used to span a 2D Fourier space in a particular
way, by sequentially adding squares whose sides are represented by
Fibonacci numbers. Likewise, we can construct a similar spiral in
3D Fourier space, the Padovan cuboid spiral, based on a sequence of
cuboids, with integer dimensions. Thus, it is possible to establish
different connections between a shell model, defined by such a recurrent
sequence of vectors $k_{n}$ and a regular Cartesian grid.

The rest of the paper is organized as follows: In Sec. \ref{sec:Goy-shell-model},
we present the GOY model revisited in order to conserve the inviscid
invariants on an arbitrary wave-number sequence. In Sec. \ref{subsec:Sequences},
we introduce various asymptotically self-similar sequences and discuss
their corresponding geometric properties. Sec. \ref{subsec:Numerical-results-I:}
compares the results between asymptotically self-similar sequences
and respective self-similar ones. An intermittency analysis is conducted
in Sec. \ref{subsec:Intermittency-analysis} demonstrating its dependence
on inter-shell spacing. Additionally, Sec. \ref{subsec:seqence=000020compond}
presents a further test of the model implemented on a wave-number
sequence that is not asymptotically self-similar, i.e. a sequence
compound in which the ratio between two consecutive term alternates.
In Sec. \ref{sec:shell-model-helical} , we introduce the helically
decomposed shell model, implemented on a generic wave-number sequence.
Finally, conclusions are presented in Sec. \ref{sec:conclusion}.

\section{Goy model on a Recurrent Sequence\protect\label{sec:Goy-shell-model}}

Let us consider a generic, shell model:

\begin{equation}
\frac{du_{n}}{dt}=i\sum_{\triangle}\Lambda_{nm\ell}u_{m}^{*}u_{\ell}^{*}-d_{n}u_{n}+f_{n}\;\text{,}\label{eq:goy}
\end{equation}
where $u_{n}$ is a complex variable representing the dynamics of
the velocity field at a given wavelength $k_{n}$, $f_{n}$ represents
external forcing and $d_{n}=\nu k_{n}^{2p}+\mu k_{n}^{-2q}$ is the
general form of (hyper/hypo)-viscosity. The forcing is usually taken
to be localized to a given shell $n=n_{f}$ or a few shells around
it, whereas the nonlinear term $\sum_{\triangle}\Lambda_{nm\ell}u_{m}^{*}u_{\ell}^{*}$
models the nonlinear interactions between shell variables, representing
a reduced subset of all the triadic interactions in Fourier space.
As a particular example that retains only local interactions the GOY
model \citep{ohkitani:89} can be written as:

\begin{equation}
\begin{aligned}\sum_{\triangle}\Lambda_{nm\ell}u_{m}^{*}u_{\ell}^{*}= & \bigg[\left(k_{n+2}+k_{n+1}\right)u_{n+2}^{*}u_{n+1}^{*}+\\
+\left(k_{n-1}-k_{n+1}\right)u_{n-1}^{*}u_{n+1}^{*} & -\left(k_{n-1}+k_{n-2}\right)u_{n-1}^{*}u_{n-2}^{*}\bigg]\;\text{,}
\end{aligned}
\label{eq:nl-goy-operator}
\end{equation}
where the coefficients are chosen such that the nonlinear term conserves
the usual quadratic quantities without any assumptions regarding the
self-similar nature of the shell variables. The conserved quantities
in question, are the two inviscid invariants that characterize 3D
Navier-Stokes turbulence, namely energy and helicity, defined as follows:

\begin{equation}
E=\sum_{n}|u_{n}|^{2},\quad H=\sum_{n}(-1)^{n}k_{n}|u_{n}|^{2}\;\text{.}\label{eq:conserv-goy}
\end{equation}

In this formulation, similar to the original GOY model, the shell
variables, carry either positive or negative helicity, depending on
whether $n$ is even or odd.

For constant shell spacing $k_{n}=k_{0}g^{n}$, the present model
can be expressed in the usual form found in literature for the GOY
model \citep{ohkitani:89}, where the nonlinear term can be written
as being proportional to:

\[
k_{n}\left(au_{n+2}^{*}u_{n+1}^{*}+\frac{b}{g}u_{n-1}^{*}u_{n+1}^{*}+\frac{c}{g^{2}}u_{n-1}^{*}u_{n-2}^{*}\right)
\]

With the parameters chosen as $a=\left(g+1\right)$, $b=\left(\frac{1}{g}-g\right)$
and $c=-\left(1+\frac{1}{g}\right)$, ensuring that $a+b+c=0$ as
required by energy conservation. In the standard GOY formulation,
the meaning of the second invariant, defined in general as $\sum_{n}\left(\frac{a}{c}\right)^{n}|u_{n}|^{2}$
can be changed by considering a separate $\epsilon$ parameter while
keeping the the inter-shell spacing (usually $g=2$) fixed, which
determines if the model is 2D like (i.e. conserving enstrophy), 3D
like (i.e. conserving helicity) or in between (see e.g. Ref. \citep{ditlevsen:1996,ditlevsen:2010}
). This means that comparing Eqn. (\ref{eq:nl-goy-operator}), to
this kind of GOY formulation (using $g$ and $\epsilon$ as separate
variables), it should be noted that in our formulation the second
conservation law is always helicity, but since we will modify the
inter\lyxadded{lorenzo}{Wed Jan 22 18:49:43 2025}{-}shell spacing,
the $\epsilon$ parameter would change accordingly so that the model
always conserves the same physical quantities, which also allows an
easier extension into arbitrary sequences.

Choosing the phase relations typical of the Sabra model \citep{lvov:98,lvov:99}
, we can, of course, also derive the Sabra version of the shell model
on our recurrent sequences. Noting that our conclusions does not depend
strongly on this choice, and since we eventually intend to extend
this approach to 2D and 3D models, such as log-lattices, and nested
polyhedra models, and since in these models, the phase relations follow
naturally from which hemisphere the particular wave-number node is
located {[}together with the condition $\mathbf{u}(\mathbf{k},t)=\mathbf{u}^{*}(-\mathbf{k},t)${]},
we only consider the GOY version in this paper.

We can then focus on the spectra obtained from shell models on these
recurrent sequences in order to characterize the effects of non-exact
self-similarity on the resulting wave-number spectra, spectral fluxes
and intermittency. One can imagine a scenario, for example, where
the breaking of the exact self-similarity symmetry of the system would
result in a different kind of multifractality. In the article \textquotedblleft Shell
Model Intermittency as Hidden Self-Similarity\textquotedblright{}
\citep{mailybaev:2022}, Mailybaev asserts that intermittency is related
to hidden self-similarity, while Aumaître et al. \citep{fauve:2024}
present intermittency as a consequence of a stationarity constraint
on energy flux, demonstrating that under this assumption, the fluctuations
of the energy flux are characterized by scaling exponents consistent
with the She-Leveque formula \citep{she:94}. The shell model presented
in this article can be implemented on a self-similar grid, an asymptotically
self-similar grid, or a non-self-similar sequence of wave numbers,
providing a potential framework to test these ideas by relaxing explicit,
built-in self-similarity of the conventional shell models.

\subsection{\protect\label{subsec:Sequences}Recurrent Sequences}

Since the primary goal of this paper is to demonstrate how the proposed
model can reproduce standard shell model features without strict self-similarity,
we follow the approach of selecting different integer sequences, representing
shell elements that lie on a regular grid and approach a constant
inter-shell spacing. In order to achieve an asymptotic ratio between
consecutive terms, we choose sequences defined through recurrence
relations. Note that the asymptotic ratio is entirely determined by
the solution of the characteristic equation derived from the recurrence
relation (e.g., $x^{2}=x+1$ for the Fibonacci sequence), rather than
by the initial terms. This approach provides us with a variety of
sequences on which the model can be implemented and compared with
their corresponding self-similar counterparts. Below, we list the
various asymptotically self-similar sequences that we have used.

Note also that, using recurrent sequences also have applications in
the context of anisotropic shell models, including multi-branch shell
models \citep{gurcan:19}, and in the recent work of Ref. \citep{vladimirova:2021}
that introduce a shell model for resonant wave interactions where
the frequencies follow the Fibonacci sequence: the resonant manifold
condition is ensured by the recurrence relation, i.e. involving resonances
between three consecutive modes. In addition to all the specific choices
presented below, which we will investigate numerically, it is worth
mentioning that given any desired inter-shell spacing $g$, it is
also always possible to construct an asymptotically self-similar sequence
of integers defined by $k_{n}=\left\lfloor k_{0}g^{n}\right\rceil $,
which rapidly converges to constant inter-shell spacing.

Another point is that the conventional GOY model with $g=2$, also
happens to be an integer sequence that one can call the ``powers
of two'' sequence. It is immediately self-similar and can be viewed
as a recurrent series with the recurrence $k_{n}=2k_{n-1}$. In this
sense, one may argue that the natural generalization of the GOY model
with $g=2$, is to recurrent series instead of non-integer $g$ values.

\subsubsection{Fibonacci and Lucas sequences}

Starting with $k_{0}=0$ and $k_{1}=1$, and constructing the rest
of the sequence using the Fibonacci recurrence relation $k_{n}=k_{n-1}+k_{n-2}$,\LyXZeroWidthSpace{}
we obtain the Fibonacci numbers. Alternatively, starting with $k_{0}=2$
and $k_{1}=1,$ with the same recurrence relation, we get the Lucas
numbers. The asymptotic inter-shell spacing of both sequences is determined
by the characteristic polynomial, whose solution gives us the asymptotic
inter-shell ratio as the golden ratio $\varphi$.

Since $\varphi$ is the maximum possible value for $k_{n}$ that forms
a triangle with $k_{n-1}$ and $k_{n-2}$, (albeit a flat one), it
can be used as a natural boundary for the shells, where the shell
centers barely interact. It also works nicely for components of a
log-lattice, which can be seen as a multi-dimensional generalizations
of shell models. Given this recurrence relation, it is also possible
to assign a geometric significance to the sequence. For instance,
the Fibonacci and Lucas spirals can be constructed simply by adding
a series of squares recursively, representing a specific partition
of the 2D Fourier space on a regular grid. More generally, using an
integer sequence guarantees that the the shell boundaries, or centers
as one may choose coincide with grid points on each of the axes. 

\subsubsection{Narayana sequence}

The Narayana sequence is defined by the third-order recurrence relation,
$k_{n}=k_{n-1}+k_{n-3}$, The first few terms are $k_{0}=k_{1}=k_{2}=1$.
thereby the asymptotic inter-shell spacing is represented by the real
root of the equation $x^{3}=x^{2}+1$, yields the supergolden ratio
$\psi\simeq1.46557\dots.$From a geometric perspective, unlike the
Fibonacci series, this sequence allows the existence of genuine triads
between shell centers. Specifically, the sides of the triangle formed
by these centers have lengths 1, $1/\psi$, and $\psi$, representing
a triangle with an obtuse angle $\theta_{kp}$ of 120 degrees (with
$\theta_{kq}=23.78^{{^\circ}}$ and $\theta_{pq}=36.22^{{^\circ}}$
using the convention $p<k<q$). The shell model implemented on this
sequence can also be interpreted as a chain of interacting modes,
connected by triads that follow this shape. Even though here we propose
a local shell model for this sequence, the generalization to log-lattices
would involve non-local interactions between the wave-numbers that
satisfy the recurrence relations.

\subsubsection{Padovan and Perrin sequences}

Considering the recurrence relation $k_{n}=k_{n-2}+k_{n-3}$, which
corresponds to the characteristic equation $x^{3}=x+1$, we obtain
sequences that asymptotically approach a spacing with the plastic
ratio $\rho\approx1.324718$, which is the real root of the characteristic
equation. Starting with $k_{0}=k_{1}=k_{2}=1$ we obtain the so-called
Padovan sequence. Alternatively starting with $k_{0}=3,\quad k_{1}=0,\quad k_{2}=2$,
one obtains the Perrin sequence. As in the case of Narayana sequence,
the log-lattice generalization would involve using the recurrence
relation at each direction, where the triadic interactions would be
between those wave-numbers that satisfy the recurrence relation for
their components.

Similar to the Fibonacci spiral, there exists a spiral construction
for the Padovan numbers, which consists of equilateral triangles whose
side lengths follow the Padovan sequence. Alternatively, in 3D Fourier
space, a spiral structure can be constructed by joining the diagonals
of the faces of successive cuboids added to an initial unit cuboid,
which is called the Padovan cuboid spiral. In this, the third dimension
of each cuboid corresponds to a successive term in the Padovan sequence,
while the other two dimensions match the length and width of the face
being extended. The sequence starts with a $1\times1\times1$ cuboid,
followed by another $1\times1\times1$ cuboid, then a $1\times1\times2$,
and next a $2\times2\times3$ cuboid. Similar to the Fibonacci spiral
for the 2D grid, the Padovan cuboid spiral can be considered as a
specific partition of the 3D Fourier space, whose the elements lie
on a regular grid, forming a chain of interacting nodes. This construction
also results in a series of triangles, each defined by two sides of
successive Padovan numbers with the angles $\theta_{kp}\approx97.04^{{^\circ}}$,
$\theta_{kq}\approx34.44^{{^\circ}}$ and $\theta_{pq}\approx48.52^{{^\circ}}$
between these sides (with the convention that $p<k<q$).

\subsubsection{Square root of the Fibonacci sequence}

As we have noted, the Fibonacci series can be used as shells in $k$-space,
but if one wants to consider them as Fourier space decimations with
$k_{n}$ representing magnitudes, they basically form only flat triangles.
In order to generate self-similar spiral-like structures, one can
instead consider a right angled triangle with two legs of length 1
and $\sqrt{\varphi}$\LyXZeroWidthSpace , so that the hypotenuse would
be $\varphi$. Repeating this procedure, we can obtain a model where
the shell spacing is the square root of the golden ratio.

In order to force such a construction on an integer sequence, we consider
the square roots of Fibonacci numbers, where a recurrence relation
can be defined as $k_{n}=\left\lfloor \sqrt{k_{n-1}^{2}+k_{n-2}^{2}}\right\rceil $,
starting with $k_{0}=1$, and $k_{1}=2$. Since the recurrence relation
involves rounding, this is more like a numerical integer sequence.
Nonetheless it is fundamentally different from simply $k_{n}=\left\lfloor k_{0}g^{n}\right\rceil $
with $g=\sqrt{\varphi}$, even though both converge to the same asymptotic
ratio.

\begin{table}[h]
\begin{tabular}{ccc}
\toprule 
 & $k_{n}$ & N\tabularnewline
\midrule 
$F_{n}$ & $k_{n}=k_{n-1}+k_{n-2}\quad\{0,1\dots\}$ & 36\tabularnewline
\midrule 
$L_{n}$ & $k_{n}=k_{n-1}+k_{n-2}\quad\{2,1,\dots$\} & 36\tabularnewline
\midrule 
$g=\varphi$ & $k_{n}=\varphi^{n}$ & 36\tabularnewline
\midrule 
$N_{n}$ & $k_{n}=k_{n-1}+k_{n-3}\quad\{1,1,1\dots\}$ & 46\tabularnewline
\midrule 
$g=\psi$ & $k_{n}=\psi^{n}$ & 46\tabularnewline
\midrule 
$P_{n}^{a}$ & $k_{n}=k_{n-2}+k_{n-3}\quad\{1,1,1,\dots\}$ & 59\tabularnewline
\midrule 
$P_{n}^{e}$ & $k_{n}=k_{n-2}+k_{n-3}\quad\{2,0,1,\dots\}$ & 59\tabularnewline
\midrule 
$g=\rho$ & $k_{n}=\rho^{n}$ & 60\tabularnewline
\midrule 
$F_{n}^{s}$ & $k_{n}=\left\lfloor \sqrt{k_{n-1}^{2}+k_{n-2}^{2}}\right\rceil \quad\{1,2,\dots\}$ & 67\tabularnewline
\midrule 
$g=\sqrt{\varphi}$ & $k_{n}=\sqrt{\varphi^{n}}$ & 67\tabularnewline
\midrule 
$C_{n}$ & $F_{n}\cup L_{n}$ & 68\tabularnewline
\bottomrule
\end{tabular}

\caption{\protect\label{tab:series-coeff}The table sums up various integer
wave-number sequences used in the numerical integrations, where the
recurrence relation and the initial values are shown, as well as the
scaling factors for their self-similar counterparts. In practice,
the first elements of the series are regularized, meaning zeros are
skipped and repetitions are avoided. The number of shells reported
in the last column are chosen in order to have roughly the same range
of wave-numbers with a constant value of $\nu=5\cdot10^{-10}$ across
all the models.}
\end{table}

\subsection{\protect\label{subsec:Numerical-results-I:} Self-similar vs. Recurrent
Sequences}

The model presented above, given $N$ shells corresponding to a sequence
of wave-numbers $k_{n}$, represents a system of $N$ coupled ordinary
differential equations (ODEs). In accordance with the literature on
the standard GOY model and the goal of estimating the intermittency
corrections, we drive the system with a constant large-scale forcing
$f_{n}=(1+i\sqrt{5})\cdot10^{-1}$ acting on the fourth shell (i.e.
\textbf{$n_{f}=4$}), and a small scale dissipation of the form $d_{n}=\nu k_{n}^{2}$,
with a small enough viscosity to dissipate energy only in the last
few shells. Although not explicitly discussed in this section, the
implementation of a forcing that is delta-correlated in time is straightforward,
and does not alter the results presented here qualitatively. In order
to integrate the system numerically we used a $4$th order Implicit-Explicit
(IMEX) solver \citep{rackauckas24,kennedy2003AdditiveRS} in Julia
\citep{julia2017}, which allows for efficient separation of the stiff
and non-stiff components of the equations.

We choose our sequences of wave-numbers so that they span seven decades
in $k$. This allows for sufficient extension of the inertial range
and is consistent with previous studies \citep{pisarenko:93,lvov:98,kadanoff1995}.
This corresponds to $N=36$ shells for the Fibonacci sequence, while
for the Padovan and the square root of Fibonacci numbers sequences,
this corresponds to $N=60$ and $N=67$ shells respectively. In contrast,
the conventional inter-shell ratio $g=2$, covers seven decades with
$N=25$ shells. The parameters for all other sequences considered
are summarized in the table \ref{tab:series-coeff}. We also chose
to fix the dissipation at $\nu=5\cdot10^{-10}$.

The system was integrated until its total energy reached a steady
state, from an initial condition of small amplitude $(10^{-8})$ white
noise, and then continued over approximately 2,000 eddy turnover times
($\tau_{E}^{-1}\sim k_{n}u_{n}$) of the largest scale after reaching
steady state. The sampling time is set to $\delta t=10^{-4}$, ensuring
a sufficiently high sampling rate to resolve the temporal dynamics
within the inertial range. All spectral quantities are averaged over
this steady state time window, as represented by $\langle\cdot\rangle$.
Recall that, since the spectral energy is defined as $E(k_{n})=k_{n}^{-1}|u_{n}|^{2}$
for shell models, a scaling of the form $\left|u_{n}\right|^{2}\sim k^{-2/3}$
is consistent with the Kolmogorov spectrum for the forward energy
cascade.

\begin{figure}[h]
\includegraphics[width=1\columnwidth]{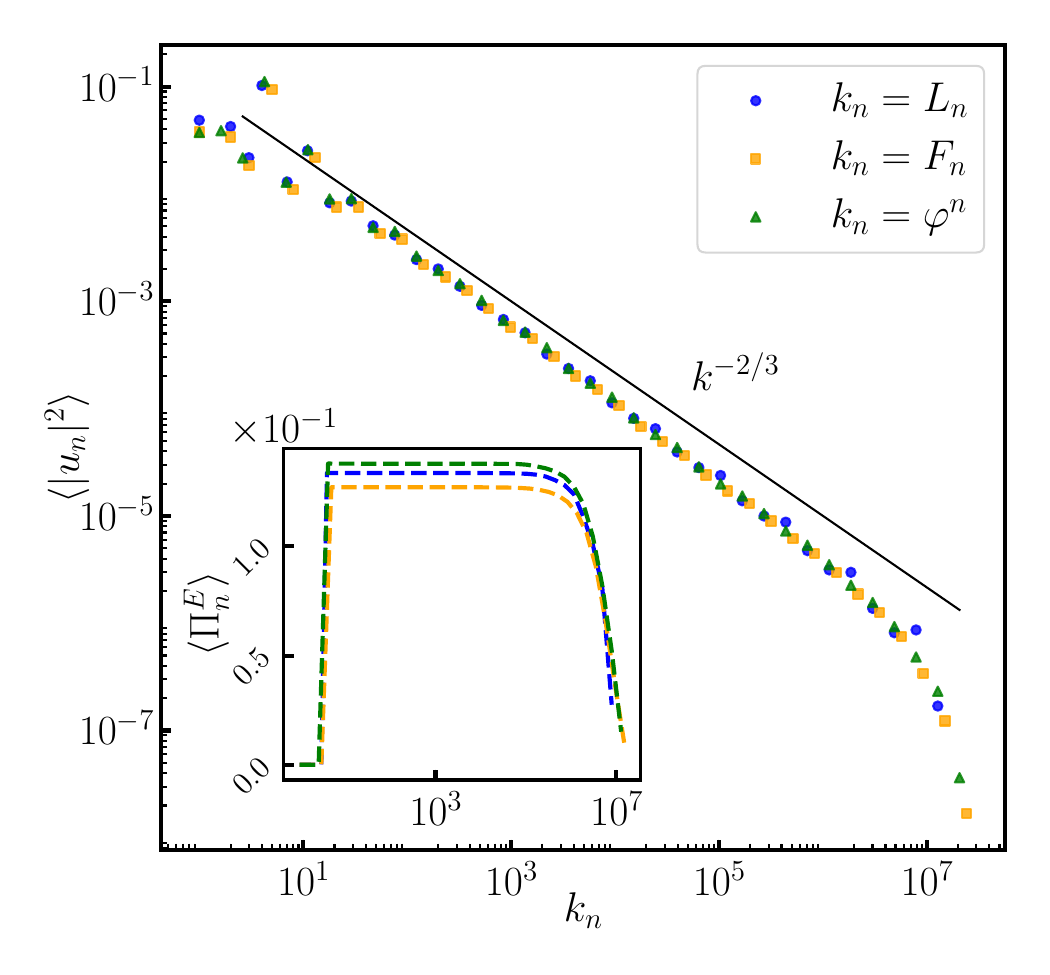}

\caption{\protect\label{fig:energy-spectra-fib}Log-log plot of the spectral
energy for each shell $\langle|u_{n}|^{2}\rangle$ as a function of
the wave-number $k_{n},$ with the wave-numbers corresponding to the
Fibonacci numbers, Lucas numbers and the respective self-similar spacing
$k_{n}\propto\varphi^{n}.$ In the inset, a semi-log plot shows the
spectral energy flux $\langle\Pi_{n}^{E}\rangle$, defined in Eq.
(\ref{eq:goy-flux-definition}), as a function of the wavenumber $k_{n}$
(log. scale).}
\end{figure}

In Figure \ref{fig:energy-spectra-fib}, the shell energy $\left|u_{n}\right|^{2}$
for the shell model implemented on the Fibonacci and Lucas sequence
is compared to the respective self-similar model (i.e. with $k_{n}=\varphi^{n}$).
We note that the spectrum exhibited by the two asymptotically self-similar
sequences is in very good agreement with the Kolmogorov scaling and
that of a GOY model with $g=\varphi$.

Following the standard GOY model definition, the spectral energy flux
at a given shell $k_{n}$ can be written as:
\begin{equation}
\begin{array}{cc}
\Pi_{n}^{E}= & 2Im(\left(k_{n+2}+k_{n+1}\right)u_{n}^{*}u_{n+1}^{*}u_{n+2}^{*}+\\
 & +\left(k_{n}+k_{n-1}\right)u_{n-1}^{*}u_{n}^{*}u_{n+1}^{*})
\end{array}\label{eq:goy-flux-definition}
\end{equation}

Note that the spectral energy flux, shown in the inset plot of Figure
\ref{fig:energy-spectra-fib}, remains perfectly constant in the inertial
range both for the GOY model with $g=\varphi$ and the shell model
on the Fibonacci and Lucas sequences. We observe that the level or
energy flux is slightly different between the three sequences. This
is because of the fact that since we force the $4$-th shell in each
case, which corresponds to a different wave-number in each sequence,
which results in different levels of energy injection.

\begin{figure}
\includegraphics[width=1\columnwidth]{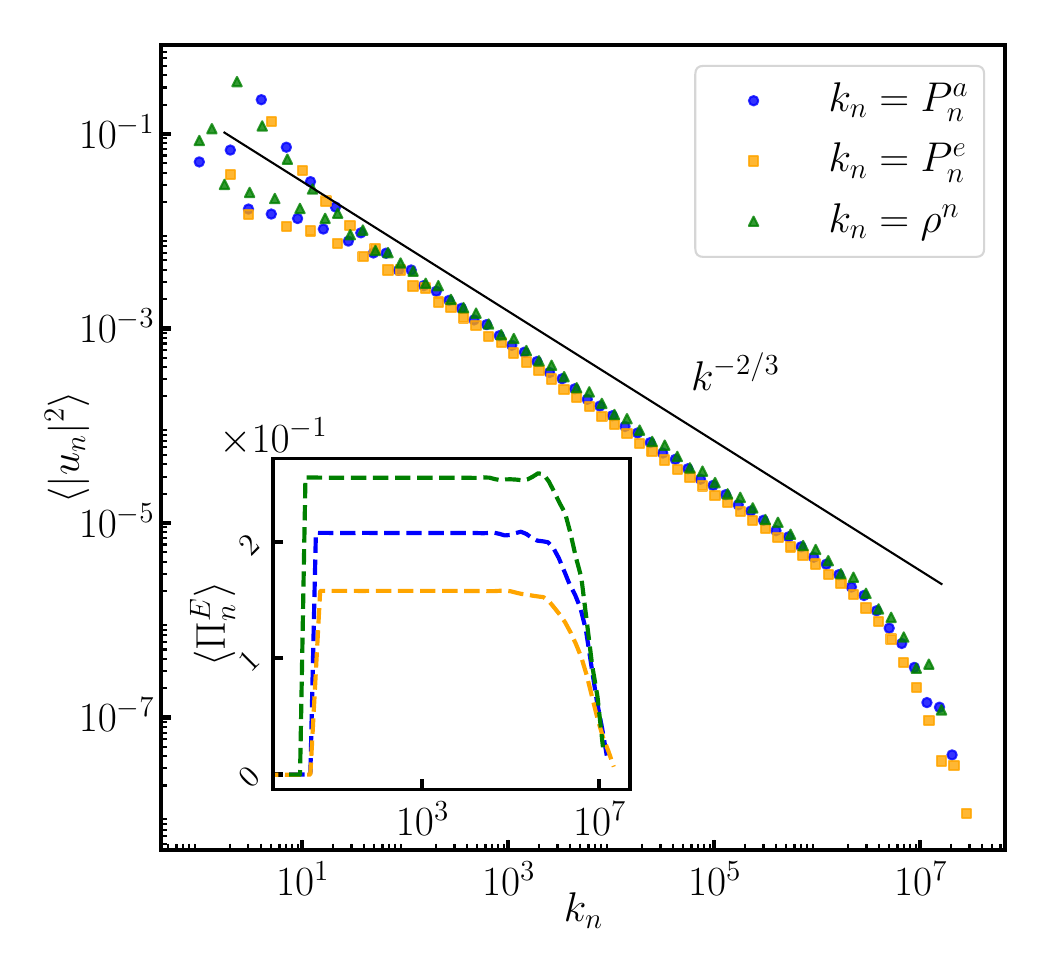}\caption{\protect\label{fig:spectra-plastic}Log-log plot of the spectral energy
for each shell $\langle|u_{n}|^{2}\rangle$ as a function of the wave-number
$k_{n},$ for a wave-number sequences corresponding to the Padovan
numbers, Perrin numbers and the respective self-similar spacing $k_{n}\propto\rho^{n}.$
The inset semi-log plot displays the spectral energy flux as a function
of the wave-number $k_{n}$ (log. scale).}
\end{figure}

As we can see in Fig. \ref{fig:spectra-plastic}, the same is true
also for the shell models with $g=\rho$ (i.e. the plastic ratio)
and the corresponding integer sequences that approach this scaling.
For the sake of brevity, we do not present the comparison between
the self-similar sequence corresponding to $g=\psi$ (i.e. the supergolden
ratio) and the Narayana sequence, which exhibit behavior similar to
the other cases presented.

\begin{figure}
\includegraphics[width=1\columnwidth]{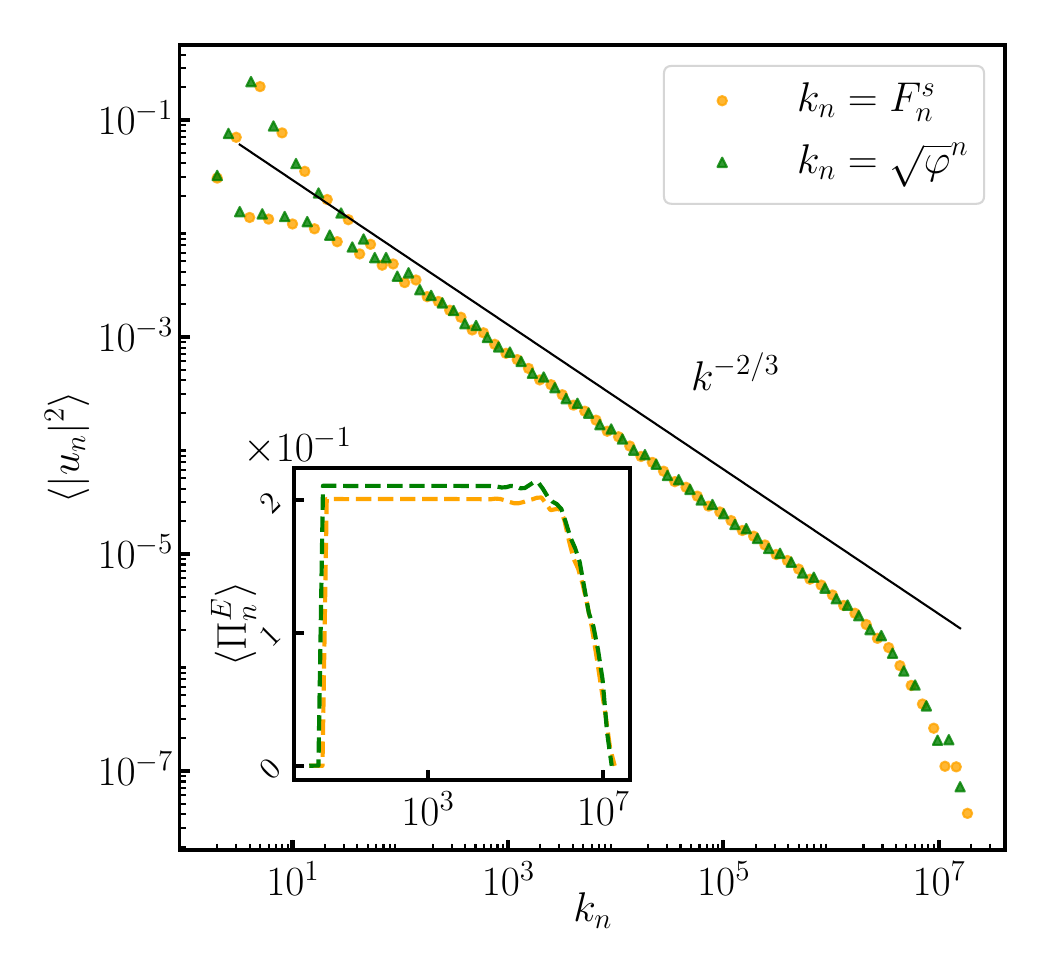}

\caption{\protect\label{fig:spectra-sqrtfib}Log-log plot of the energy spectra
for the square root of the Fibonacci sequence and the corresponding
self-similar sequence. The inset semi-log plot displays the spectral
energy flux as a function of the wave-number $k_{n}$ (log. scale).}
\end{figure}

The same comparison can be repeated between the GOY model with $g=\sqrt{\varphi}$,
and the associated sequence made of square roots of Fibonacci numbers
rounded to the nearest integers. Again, the conclusion is that the
two models behave very similar to one another, as can be seen from
the plot of the spectral energy (Fig. \ref{fig:spectra-sqrtfib}).
One notable observation is that the amplitude of the oscillations
at the forcing scale seems to be larger, for smaller spacing, i.e.
Plastic ratio and square root of golden ratio, compared to the Fibonacci
case, before eventually reaching a specific scaling corresponding
to a constant flux in the inertial range, as illustrated in the Figure
\ref{fig:spectra-sqrtfib}. At the Kolmogorov scale, for smaller spacing,
the flux presents some oscillations compared to the Fibonacci case.
We also observe that the energy spectrum starts to shows slight deviations
from the Kolmogorov scaling, suggesting the presence of higher-order
corrections.


\subsection{\protect\label{subsec:Intermittency-analysis}Intermittency analysis}

It is well-known that the higher order structure functions computed
from shell models (i.e. $S_{p}\equiv\left\langle \left|u_{n}\right|^{p}\right\rangle $),
with the usual value of $g=2$, exhibit deviations from the Kolmogorov
scaling of $S_{p}\propto k_{n}^{-p/3}$, demonstrating intermittency
\citep{jensen:91,pisarenko:93,kadanoff1995,de-wit:2024:}, somehow
consistent with the intermittency in realistic flows.

\begin{figure*}
\includegraphics[width=1\textwidth]{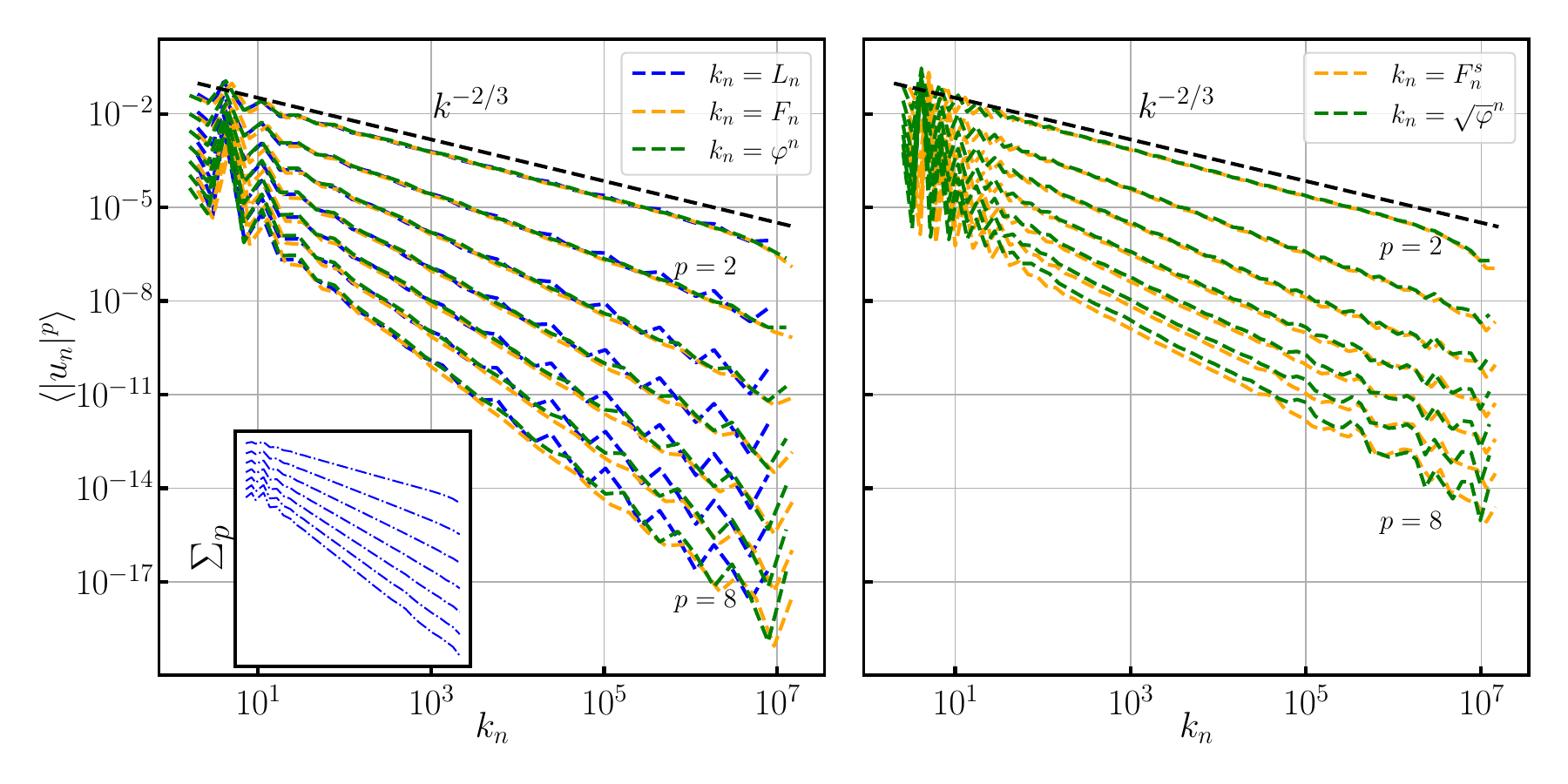}

\caption{\protect\label{fig:structuref_fib-vs-sqrt}Log-log plot of the structure
functions of order $p$ (from two to eight, top to bottom), defined
as $S_{p}(k_{n})=\langle|u(k_{n})|^{p}\rangle$ as a function of the
wave-number $k_{n}$, for both the self similar spaced model and the
asymptotically self similar one in two different spacing considered,
the golden ratio and the square root of the golden ratio. The inset
plot indicates the structure function after eliminating the periodic
three oscillations, showing the scaling of the quantity $\Sigma_{p}$.}
\end{figure*}
It is also well-known that the GOY model exhibits a static solution
consisting of cyclic oscillations involving three consecutive shells
with respect to the shell index (as shown in \citep{biferale:95}),
which overlap with the scaling of the structure function, making the
estimation of scaling laws less accurate. It is possible to filter
out these oscillations \citep{kadanoff1995}, by studying the scaling
behavior of the quantity constructed in terms of three point correlations
as follows:

\[
\Sigma_{n,p}=\left\langle \left|Im\left(u_{n-1}^{*}u_{n}^{*}u_{n+1}^{*}\right)\right|^{p/3}\right\rangle 
\]

We can characterize the intermittency the same way in our model, and
compare it with the intermittency in the standard GOY case. Note that
this is a more stringent test, and a potentially more interesting
comparison since the Fibonacci sequence shell model breaks the exact
self-similar structure of the GOY model. As can be seen in the inset
plot of Figure \ref{fig:structuref_fib-vs-sqrt}, the scaling of $\Sigma_{p}$
is not affected by these three-point oscillations, therefore it allows
for a slightly more accurate estimation of the scaling exponents.

\begin{figure}
\includegraphics[width=1\columnwidth]{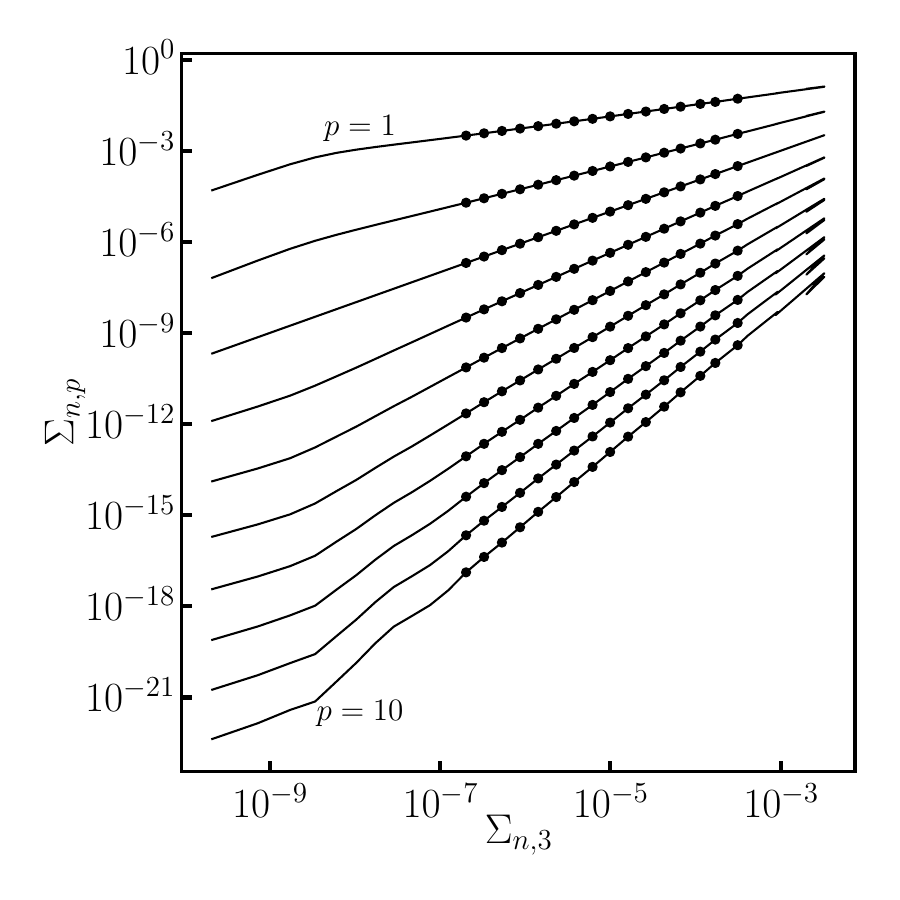}

\caption{\protect\label{fig:EES-procedure}In the figure, for the Fibonacci
sequence case, we illustrate the procedure for estimating the scaling
exponents $\xi(p)$ following the ESS method. The scalings of the
quantity $\Sigma_{p}$ are plotted as a function of $\Sigma_{3}.$
The structure function $p=1,2,\dots,10$ are displayed from top to
bottom. The dots represent the set of points used for the fitting
procedure, corresponding to approximately four decades in the inertial
range, which aligns with the constant flux window.}
\end{figure}

The scaling exponent $\xi(p)$ of the structure function of order
$p$ can also be estimated via an extended self similarity (ESS) procedure
\citep{benzi93} that allows an improved determination of the intermittency,
due to the reduction of eventual subdominant contributions to the
scaling: through a linear fit of the ordinate as a function of the
abscissa in a log-log plot of the structure function $S_{p}$ or $\Sigma_{p}$
as a function of the third order structure function $S_{p=3}$ ($\Sigma_{p=3}$
), assuming a power law scaling of the form $S_{p}(k_{n})\sim S_{3}^{-\xi(p)}$
($\Sigma_{p}(k_{n})\sim\Sigma_{3}^{-\xi(p)}$) in the inertial range,
selecting a range of shells that correspond to a perfectly constant
flux, as can be seen for different values of $p$ in Figure \ref{fig:EES-procedure}.

As previously noted, spacings smaller than the golden ratio, where
actual triadic interactions are possible with the shell centers, such
as the square root of golden ratio spacing or the plastic ratio spacing,
be it on an exactly or asymptotically self-similar model, appear to
exhibit behavior that does not align perfectly with the Kolmogorov
scaling, suggesting higher-order corrections. First, the quantities
$\left\langle \left|u_{n}\right|^{p}\right\rangle $ are shown in
Figure \ref{fig:structuref_fib-vs-sqrt} for shell models with golden
ratio and the square root of golden ratio spacings. This demonstrates
that the scaling behavior for these two different types of spacing
is significantly different, while indicating no substantial differences
from a given integer sequence and their corresponding self-similar
counterpart.

\begin{figure}
\includegraphics[width=1\columnwidth]{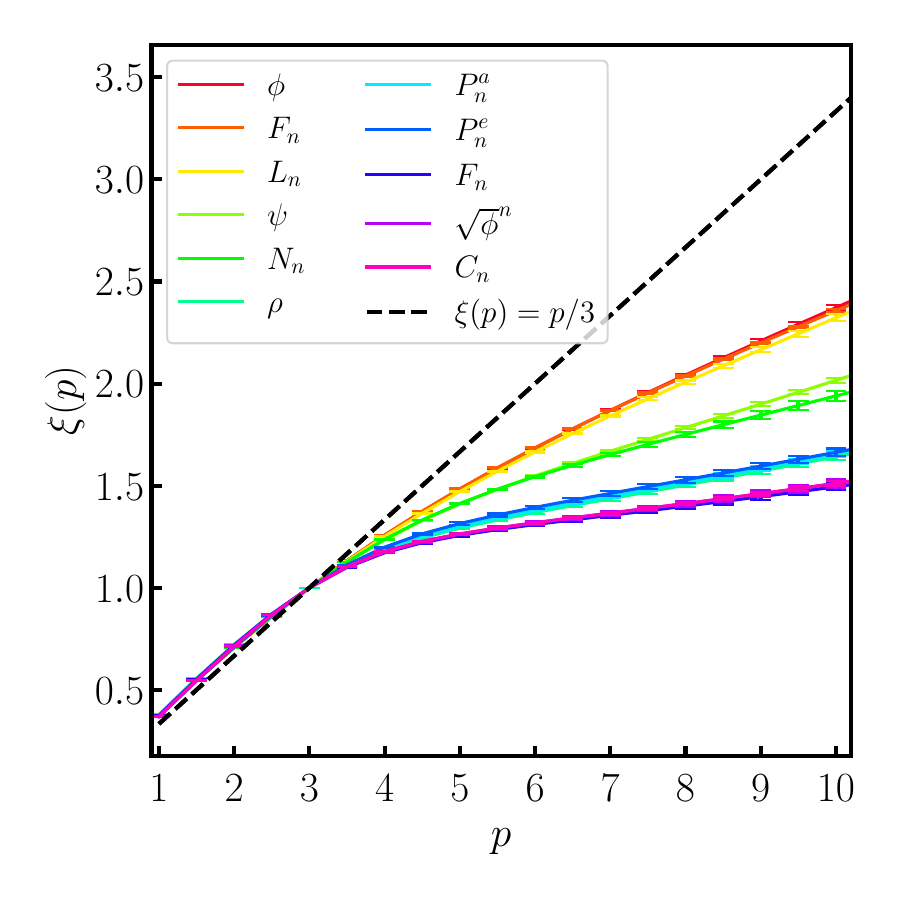}

\caption{\protect\label{fig:intermittency}Plot of the scaling exponent $\xi(p)$
as a function of the order of the structure function estimated via
the ESS procedure for all the sequences and corresponding self similar
sequences studied, the parameters utilized corresponds to what displayed
in table \ref{tab:series-coeff}. The dashed line correspond to the
Kolmogorov scaling $\xi(p)=p/3$.}
\end{figure}

The scaling factors $\xi(p)$ that are estimated through an ESS procedure
are plotted in Figure \ref{fig:intermittency} as a function of $p$,
the order of the structure function, in order to quantify the intermittency
corrections for all the sequences that have been studied. As usual,
the deviation from the linear scaling serves as a measure of intermittency.
Note that we have also performed the simpler procedure of fitting
the structure functions directly and the results were essentially
the same.

First, we note that all the curves clearly exhibit very high deviations
from what would be a mono-fractal scaling, which may be interpreted
as extremely high levels of intermittency. In addition, for each family
of curves, i.e., for every asymptotic scaling ratio, we can observe
that the intermittency of the integer sequences follows closely the
GOY version with that scaling, within the error bars of the fitting
procedure and the slight differences in injection and dissipation.
However, we note an interesting, and somewhat unexpected result: for
each family of series, intermittency follows a different trend, suggesting
that smaller shell spacing corresponds to larger intermittency corrections.

Additionally, we perform an analysis of intermittency over time, following
the approach outlined in Ref. \citep{mitra-pandit:2004,pandit:08,ray:08},
which we do not present here, but which shows similar trends. We find
that the presented results are relatively robust with clear scaling
laws demonstrated by the structure functions, having tested three
different series for each asymptotic scaling factors. These results
are non-trivial and quite novel, considering that Sabra and shell
models have traditionally been explored numerically with an inter-shell
spacing of $g=2,\varphi,$ and that smaller shell spacing have been
examined only in a few examples (\citep{kadanoff1995,benzi:1996,benzi:96}).

As additional evidence that the proposed form can reproduce standard
shell model features without self-similarity of shell wave-numbers,
we also considered the Sabra version of the model using recurrent
sequences and corresponding self-similar ones. We observe the same
qualitative behavior: self-similar and recurrent sequences exhibit
same energy and flux scalings and comparable levels of intermittency,
and decreasing the inter-shell ratio results in an increase of the
intermittency corrections, even though this latter effect seems to
be less pronounced in the Sabra case.

An in-depth characterization of the intermittency dependence on varying
inter-shell spacing of standard shell models is beyond the scope of
the present paper, whose primary goal is to demonstrate the robustness
of the implementation of shell models on recurrent sequences, leaving
a more thorough investigation of the former for a future publication.

\subsection{\protect\label{subsec:seqence=000020compond}A two sequence compound
example}

In this section, we demonstrate the behavior of the proposed model
on a two-sequence compound constructed by merging two sequences. As
a particular example, we consider the compound formed by the union
of the Fibonacci and Lucas numbers. These two sequences share the
same recurrence relations, and the same asymptotic scaling factor,
the golden ratio. The Fibonacci and Lucas sequences are complementary,
and one can construct similar spirals that rotate in co and counter-clockwise
directions from them respectively, which for large numbers, approximate
golden spirals common also in pattern formation in plants such as
phyllotaxis \citep{newell:08}. In this sense, it is a compound sequence:
and the two sequences asymptotically share the same self-similar spacing,
i.e. the golden number, they begin with slightly different numbers
but grow with the same scaling factor, thus not overlapping but forming
a new sequence that can be expressed as:

\[
\left\{ \dots F_{n},L_{n-1},F_{n+1},L_{n},F_{n+2},L_{n+2},\dots\right\} 
\]

Consequently, the shell model constructed in the asymptotic limit
does not converge to a constant inter-shell ratio. Instead, the inter-shell
ratio alternates between two values:

\[
\frac{k_{n+1}}{k_{n}}=\begin{cases}
\frac{F_{n+2}}{L_{n}}=\frac{\varphi^{2}}{\sqrt{5}}\simeq1.171\\
\frac{L_{n}}{F_{n+1}}=\frac{\sqrt{5}}{\varphi}\simeq1.382
\end{cases}
\]

Considering two consecutive shells lumped together as one, the inter-shell
spacing would approach the golden ratio. Note also that there are
two kinds of triads in the compound. The first triad with $L_{n}$
as the central node is an acute triangle with the asymptotic angles
$\theta_{kq}=37.92{^\circ}$, $\theta_{kp}=83.94{^\circ}$ and $\theta_{pq}=58.14{^\circ}$,
and the second one with $F_{n}$ as the central node is an obtuse
one with the angles $\theta_{kq}=37.92{^\circ}$, $\theta_{kp}=96.06{^\circ}$
and $\theta_{pq}=46.02{^\circ}$.

We note that this type of construction has emerged in previous results
\citep{gurcan:17}, which developed a model based on a decimation
of Fourier space using the vertices of nested polyhedra, allowing
for the existence of genuine triad conditions among three modes belonging
to three consecutive ``shells'': the compound in this case was formed
by a sequence of consecutive dodecahedron-icosahedron structures,
scaled by the golden ratio. In this type of decimation, the inter-shell
ratio alternates due to the differing scaling factors between the
icosahedron and the dodecahedron.

Remarkably, in this particular type of Fourier space decimation, the
nested polyhedra model shows no sign of intermittency. Therefore,
implementing a compound sequence in a 1D chain is intriguing, especially
given the trends observed in the previous section on asymptotically
self-similar sequences: smaller spacing tends to indicate an increase
in intermittency corrections. \emph{A priori}, it is unclear whether
the model will exhibit a level of intermittency comparable to that
observed in the Fibonacci sequence or similar to the one found in
sequences with smaller inter-shell spacing, or have no intermittency
as in the case of nested polyhedra models.

\begin{figure}
\includegraphics[width=1\columnwidth]{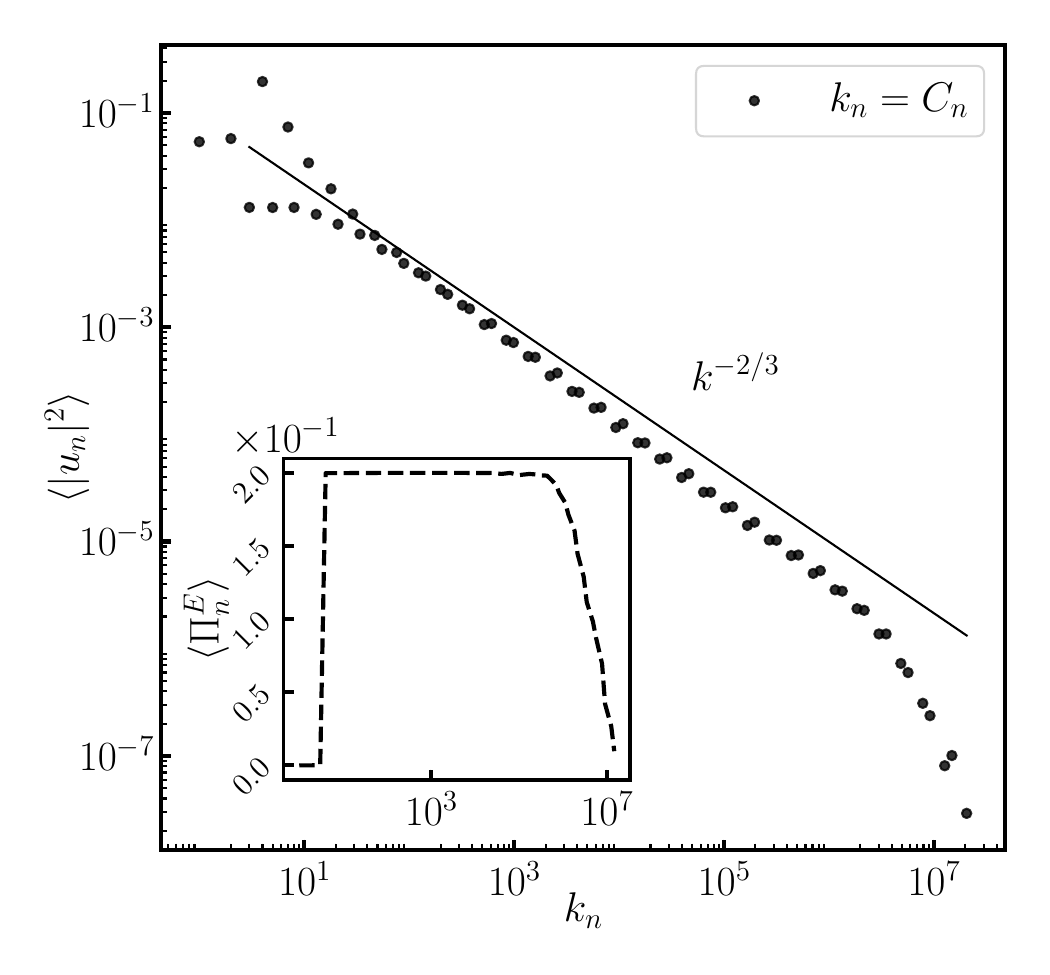}

\caption{\protect\label{fig:spectra-flux-compound}Log-log plot of the energy
spectra as function of the wave-number $k_{n}$ for the compound sequence
case. The inset semi-log plot represents the spectral energy flux.}
\end{figure}

As shown in Figure \ref{fig:spectra-flux-compound}, the model implemented
on the compound sequence reproduces results consistent with Kolmogorov
scaling. The spectrum shows the same features observed in other cases,
such as oscillations at the forcing scale and smaller in the inertial
range. These oscillations are analogous to the period-three oscillations
of the standard GOY model. However, we note that the 1D map that generates
the stationary inviscid solution \citep{biferale:95} is modified
by the fact that the shell wave-numbers are not asymptotically regular.
The energy flux remains remarkably constant in the inertial range,
which also serves to demonstrate that the proposed nonlinear term
(\ref{eq:nl-goy-operator}) generalize the usual GOY model on arbitrary
sequences of wave-numbers.

Regarding the intermittency, it is interesting to note that the scaling
of $\xi(p)$, as shown in Figure \ref{fig:intermittency}, displays
a scaling behavior very similar to that observed for the two series
with a spacing of the square root of the golden ratio (exactly or
asymptotically). Actually, if we take the geometric mean of the two
inter-shell ratios, we obtain the square root of the golden ratio.
Therefore, we hypothesize that, also in the compound sequence case,
the level of intermittency is determined merely by the spacing. Specifically,
in the sense that it is the average spacing of the sequence of shells
that determines the level of intermittency. It is puzzling however,
how one goes from that, to ``no intermittency'' in the case of the
nested polyhedra model, suggesting that the phase dynamics on the
constrained $1D$ chain structure of the shell model has something
to do with the observed dependency of the intermittency to shell spacing.
Again a detailed exploration of this observation, is left to a future
publication.

Note finally that the compound, which we argue to correspond to an
integer series version of the nested polyhedra with its alternating
triads, can also be used to construct a recurrent series version of
the log-lattice. Given that $L_{n}=F_{n-1}+F_{n+1}$ and $5F_{n}=L_{n+1}-L_{n-1}$,
this results in more triads per node (i.e. 18 for Fibonacci, to 32
and 50 for the compound), and more ``non-local'' interactions.

\section{Helical Shell models on a Recurrent Sequence \protect\label{sec:shell-model-helical}}

In the previous section, a generalization of the conventional GOY
model to generic wave-number sequences, which are in practice obtained
from recurrence relations and have asymptotically self-similar structures
is proposed. It is demonstrated that shell models on such recurrent
sequences exhibit standard features of conventional shell models such
as power law spectra, constant spectral flux, and intermittency. A
particular form of the interaction coefficients were used for this
purpose, which guarantees conservation of energy and helicity as in
the original equations.

However, in line with the usual convention in shell models, the helicity
in this picture was defined as $H=\sum_{n}(-1)^{n}k_{n}|u_{n}|^{2}$,
which allows one to enforce the conservation of a non-positive defined
invariant that scales as helicity, but does not in fact correspond
to it in any meaningful sense. Therefore, if we want to to account
for the complex role played by helicity in turbulent cascade \citep{waleffe:93,biferale:13,alexakis2017,alexakis:18,sahoo:18},
it is necessary to generalize the model to include a conserved non-positive
defined invariant that is closer to the usual definition of helicity,
as derived from the Navier-Stokes equations.

In order to achieve this, we can use the so called helical decomposition
\citep{waleffe:92}, and following De Pietro \emph{et al.} \citep{depietro:15},
write a set of arbitrarily elongated shell models, but on a recurrent
sequence in accordance with the models considered above. The resulting
model evolves two shell variables $u_{n}^{+}$ and $u_{n}^{-}$, with
respectively positive and negative helicities and allows us to define
energy and helicity as follows:

\begin{equation}
E=\sum_{n}|u_{n}^{+}|^{2}+|u_{n}^{-}|^{2},\quad H=\sum_{n}k_{n}(|u_{n}^{+}|^{2}-|u_{n}^{-}|^{2})
\end{equation}

The general form of the shell model, omitting the dissipative and
forcing terms, with arbitrary elongation can be written in compact
form as:

\[
\frac{du_{n}^{s_{0}}}{dt}=s_{0}\left(a_{n}u_{n+\ell+m}^{s_{3}*}u_{n+\ell}^{s_{1}*}+b_{n}u_{n-\ell}^{s_{1}*}u_{n+m}^{s_{2}*}+c_{n}u_{n-m}^{s_{2}*}u_{n-\ell-m}^{s_{3}*}\right)
\]
with

\begin{align*}
 & a_{n}=Q\frac{s_{1}}{s_{0}}\left(\frac{s_{2}}{s_{0}}k_{n+\ell+m}-k_{n+\ell}\right)\\
 & b_{n}=Q\left(\frac{s_{1}}{s_{0}}k_{n-\ell}-\frac{s_{2}}{s_{0}}k_{n+m}\right)\\
 & c_{n}=Q\frac{s_{2}}{s_{0}}\left(k_{n-m}-\frac{s_{1}}{s_{0}}k_{n-\ell-m}\right)
\end{align*}
$s_{3}\equiv s_{0}s_{1}s_{2},$ and where $Q=Q_{\ell m}\left(s_{1}/s_{0},s_{2}/s_{0}\right)$
is the geometric factor of the triad class that is considered. The
four triad classes can be defined as the $4$ possible combinations
of the signs of $s_{1}/s_{0}$ and $s_{2}/s_{0}$. In practice for
a given set of wave-vectors $p<k<q$ which in this case corresponds
to $p=k_{n-\ell}$ and $q=k_{n+m}$, it can be written as:
\begin{align*}
Q & \equiv Q_{kpq}^{s_{0}s_{1}s_{2}}=Q\left(\frac{p}{k},\frac{s_{1}}{s_{0}};\frac{q}{k},\frac{s_{2}}{s_{0}}\right)\\
 & =\frac{s_{1}}{s_{0}}\frac{s_{2}}{s_{0}}\sin\left(\alpha+\beta\right)\left(1+\frac{s_{1}}{s_{0}}\frac{p}{k}+\frac{s_{2}}{s_{0}}\frac{q}{k}\right)
\end{align*}
where $\alpha$ and $\beta$ are the angles between $k$ and $p$
and $k$ and $q$ respectively, which can be written as usual as:
\begin{align*}
\alpha & =\arccos\left(\frac{q^{2}-k^{2}-p^{2}}{2kp}\right)\\
\beta & =\arccos\left(\frac{p^{2}-k^{2}-q^{2}}{2qk}\right)
\end{align*}
Notice that, by choosing $g=2$, $\ell=1$, $m=1$, we obtain the
simple helical shell model discussed by Benzi \emph{et al.} \citep{benzi:96}.
However the general form of the model allows all kinds of models with
non-local interactions.

It can be shown that the proposed form of the helical shell model,
conserves energy and helicity. One can also put all the $4$ classes
of interactions together by considering a sum over triad classes on
the right hand side, with the corresponding geometric factors for
each class as discussed in Ref. \citep{rathmann} for local interactions.
Note that even though, on a recurrent sequence, the actual geometric
factors are not exactly the same, and in the above form we take them
to be the same, the energy and helicity are still properly conserved.
This is non-trivial if one thinks in terms of conventional shell models
and self-similarity. However it becomes trivial if one thinks in terms
of a chain of triads, each triad conserves energy separately regardless
of the exact geometric factor, so as long as one keeps all the nodes
of each triad that is considered and all the geometric factors of
a triad remain the same (it doesn't matter if the actual value of
the geometric factor may be a bit off), all the conservation laws
of the original system will continue to be respected.

Note that here, we presented a general helical shell model based on
the symmetries of the helically decomposed turbulence. Below we will
consider a particular class and range of interactions and show that
in a particular configuration, such a model can give rise to inverse
cascade. However, in order to make connection to the GOY model, consider
alternatively the class 4, which is defined by the relations $s_{1}/s_{0}=s_{2}/s_{0}=-1$.
This means that the middle wave-number is of opposite chirality to
the other two, and therefore we can setup two chains of nodes with
alternating chiralities that never interact. In the first chain, the
even nodes would $+$ and odd ones would be $-$, in the second chain
the odd nodes would be $+$ and even ones would be $-$. This explains
the usual interpretation of the helicities carried by shell variables
in the GOY model.

\subsection{Inverse cascade on a recurrent sequence}

In this section, we present the numerical implementation of the helical
shell model on a recurrent wave-number sequence, in order to demonstrate
that the agreement between the recurrent and self-similar shell model
is not model dependent.

Helically decomposed shells model have been studied in the case of
local interactions, both in a 1D chain of interacting modes and in
hierarchical trees in Ref. \citep{benzi:96,benzi:1997}. It is also
interesting to note, in accordance with the results of the previous
section, that in Ref. \citep{benzi:1996}, a particular class of helical
shell model shows that a decrease in the inter-shell spacing leads
to a reduction in intermittency corrections.

More recently non-local interactions have been included in the shell
model formulation in Ref. \citep{depietro:15,rathmann}. The nontrivial
aspect of including non-local couplings, arises from a class of heterochiral
interactions, namely $s_{1}/s_{0}=-1,s_{2}/s_{0}=+1$, which for sufficiently
elongated triads drive an inverse cascade. The transition from direct
to inverse cascade was first predicted by Waleffe \citep{waleffe:92,waleffe:93}
at the threshold value $p/k<0.278$ (assuming $p<k<q$ ) in the shape
of the triad interaction, and numerically confirmed by De Pietro \emph{et
al.} \citep{depietro:15}. The inverse cascade the helical shell model
represents an anomaly, as it is well known that constructing a shell
model that conserves the inviscid invariants of 2D turbulence, namely
energy and enstrophy, the inverse cascade range is typically dominated
by an equipartition spectrum as observed in Ref. \citep{aurell:94b,ditlevsen:1996,aurell:94}.
The interplay between equilibrium and cascades solutions as been further
investigated by \citet{ditlevsen:1996,gilbert:2002,tom:2017}.

We implement the helical shell model formulation on Lucas and Fibonacci
sequences and the corresponding self similar spaced model, in a numerical
setup capable of demonstrating an inverse energy cascade. Thus, we
cover seven decades with $N=36$ shells, and in order to drive an
inverse cascade $s_{1}/s_{0}=-1,s_{2}/s_{0}=+1$ and , $l=3$ , $m=1$,
resulting asymptotically in $p/k\approx1/\varphi^{3}\approx0.236$,
which satisfies the inequality predicted by Waleffe.

The system is driven by small scale random forcing acting on two consecutive
shells, $n_{f}=32,33$. The forcing amplitude is unbalanced between
the shells carrying positive and negative helicities, with $f^{+}=f$,
$f^{-}=1/2f$, where we set $f=0.8$. The dissipative term $d_{n}=\nu k_{n}^{2p}+\mu k_{n}^{-2q}$
, has the form of a small scale dissipation, with $\nu=4\cdot10^{-12}$,
$p=1$, and a large scale hypo-viscosity with $\mu=1,$ $q=2$ , in
order to prevent energy accumulation on the first shells.

\begin{figure}
\includegraphics[width=1\columnwidth]{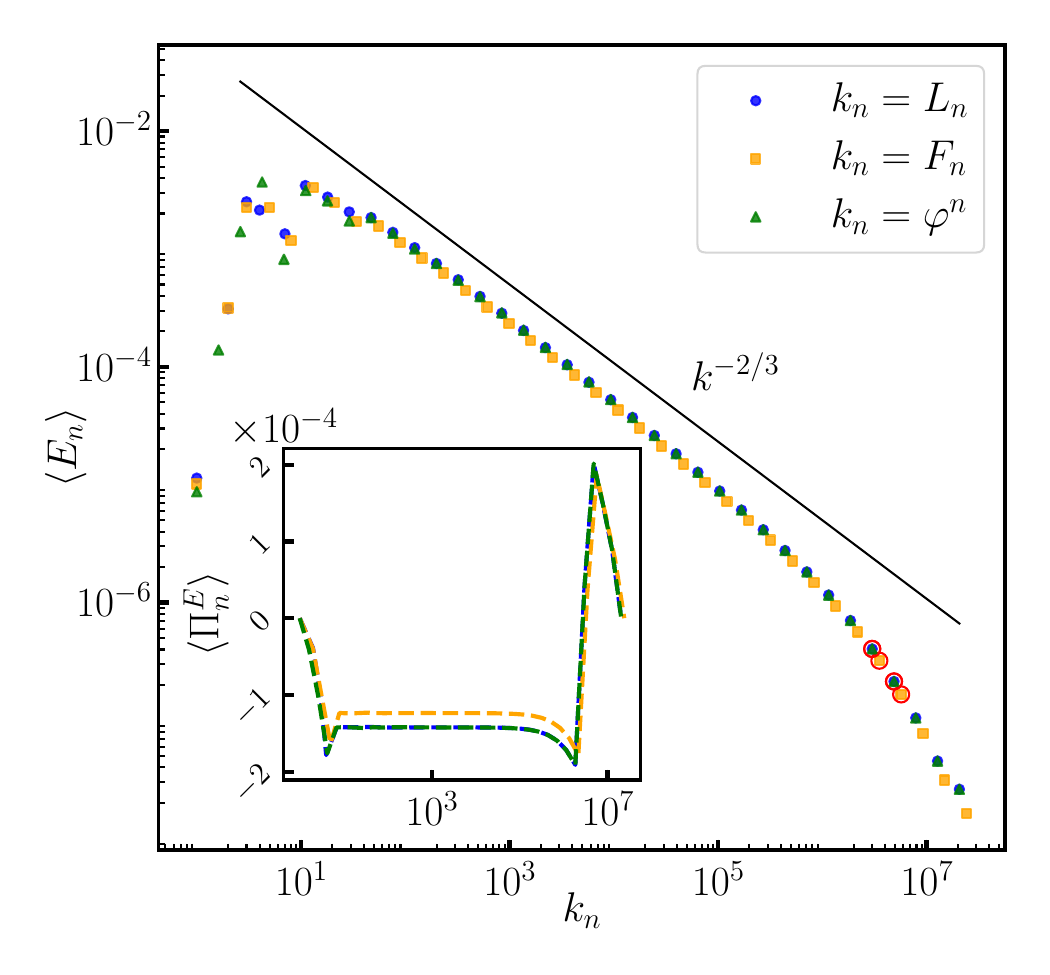}

\caption{\protect\label{fig:energy-spectrum-inverse-helical}Log-log plot of
the spectral energy for each shell $\langle|u_{n}^{+}|^{2}+|u_{n}^{-}|^{2}\rangle$
as a function of the wave-number $k_{n},$ for a wave-number sequences
corresponding to the Fibonacci numbers, Lucas numbers and the respective
self-similar spacing $k_{n}\propto\varphi^{n}.$ For each sequence
the two shells where the forcing acts are marked with a red circle.
In the inset, a semi-log plot shows the spectral energy flux $\langle\Pi_{n}^{E}\rangle$
due to nonlinear terms defined in Eq. (\ref{eq:helical-flux}), as
a function of the wave-number $k_{n}$ (log. scale).}
\end{figure}

As shown in Fig. \ref{fig:energy-spectrum-inverse-helical}, by forcing
the system at small scales an inverse energy cascade develops for
sufficiently elongated triads. We observe how the energy spectrum
is in good agreement with the inverse energy cascade prediction for
both Lucas and Fibonacci sequences and the corresponding self-similar
sequence, with three models overlapping in the inertial range. However
the energy of the first few shells differ, mainly due to the role
played by hypoviscosity : variations in the first wave numbers between
the sequences, result in different damping effects.

The flux due to nonlinear terms across the $n$-th shell can be defined
by as:

\begin{equation}
\Pi_{n}^{E}=\sum_{j=1}^{n}2Im\left(u_{j}^{+*}\left.\frac{d}{dt}u_{j}^{+*}\right|_{nl}+u_{j}^{-*}\left.\frac{d}{dt}u_{j}^{-*}\right|_{nl}\right)\;\text{,}\label{eq:helical-flux}
\end{equation}
where the subscript $nl$ denote the nonlinear terms. The spectral
fluxes for the inverse cascade on the different sequences that are
considered are shown in the inset plot of Fig. \ref{fig:energy-spectrum-inverse-helical}.
We note that the negative flux remains perfectly constant within the
inertial range, as obtained for the GOY model implementation. The
magnitude of the flux differs slightly between the Fibonacci and the
other two sequences coincide, as the energy injection varies due to
the differences in the shell wave numbers: asymptotically, the wave
numbers $k_{n}=\varphi^{n}$ align with the Lucas sequence, but is
shifted with respect to the Fibonacci sequence, even though all sequences
converge to the same inter-shell spacing. The negative flux exhibits
a cusp in all three models around the fourth shell, where do to the
elongation of the triad $l=3$ the term $n-l$ reach the boundary. 

We conclude that, also for the helically decomposed shell model, the
behavior on the recurrent integer sequences shows no significant differences
compared to that on the corresponding self-similar scaling.

\section{\protect\label{sec:conclusion}conclusion}

Considering shell models as prototypes or building blocks of a whole
host of reduced models, we propose a natural generalization of the
GOY model (i.e. with $g=2$, hence on a recurrent sequence of powers
of two), into \emph{other} recurrent sequences, including Fibonacci,
Lucas, Padovan and Narayana series and their combinations, which can
be constructed from simple additive recurrence relations (such as
$k_{n}=k_{n-1}+k_{n-2}$ ). Such series tend to generate integer numbers
that are asymptotically self-similar (i.e. $k_{n+1}/k_{n}\approx k_{n}/k_{n-1}\approx g$
for $n\gg1$).

Considering a particular form of the interaction coefficients, a model
can be constructed that respects the conservation laws of the original
system regardless of the shell spacing. This allows the implementation
of shell models on recurrent sequences characterized by different
asymptotic inter-shell spacing, and their comparison to the conventional
shell models with a constant $g$ equal to the asymptotic ratio showed
no substantial differences. In particular, the intermittency analysis
on various sequences show how the deviation from the Kolmogorov scaling
are determined by the inter shell spacing rather than whether or not
the model is on a recurrent sequence or have constant logarithmic
spacing. It was observed in particular that smaller spacings lead
to higher apparent intermittency as defined in the usual way. Suggesting
that burst-like solutions, to which the GOY model is prone in the
continuum limit \citep{Andersen:00}, may be an explanation of the
apparent high intermittency with small inter-shell spacing.

The particular case of the compound sequence, clearly demonstrates
how the proposed model is robust, showing a spectral behavior in accordance
with the regular GOY model, even though the inter-shell spacing in
such a model alternates even for large $n$. The intermittency corrections
are comparable to those obtained for a self-similar model with a spacing
equal to the geometric mean of the two alternating spacings, suggesting
again that in a constrained 1D chain of interacting modes, phase dynamics
approaching the continuum limit is the key mechanism of intermittency
in shell models.

In addition to these additive recurrence relations, we have also considered
what we call the square root of Fibonacci numbers sequence, which
follows a recurrence that one obtains by rounding the squares of the
sums of the last two elements to the nearest integer (i.e. $k_{n}=\left\lfloor \sqrt{k_{n-1}^{2}+k_{n-2}^{2}}\right\rceil $).
Like the other series with $g<\varphi$, such series allow actual
triads, and hence can be used for example to construct logarithmically
discretized models, where the discretization in the magnitude would
be constructed from the recurrence relations, while the angular discretization
would be linear.

Finally, with the same objective, we presented the formulation of
a helically decomposed shell model on a recurrent wave-number sequence
with generic elongations and helical interactions classes, generalizing
the work of \citep{rathmann,depietro:15}. We tested this construction
by demonstrating an inverse energy cascade on a recurrent sequence,
which validates the previous observation that shell models on recurrent
sequences display the same characteristics as standard shell models.

It is also worth re-iterating that the primary motivation of this
work has been to make the connection between regular and logarithmic
discretization used in turbulence modeling. By showing that this can
be achieved by recurrent sequences in shell models, paves the way
to applying the same principle to other logarithmically discretized
systems, such as log-lattices, LDM models, nested-polyhedra models
etc. It can be shown for example that using the Fibonacci or the Lucas
sequences, a log-lattice formulation would result in 18 triads per
wave-number element (at least two of which are flat triads), whereas
if one used the compound sequence made up of both Lucas and Fibonacci
numbers, the number of triads goes up to 32 for Fibonacci nodes, and
50 for Lucas nodes. Note that recurrent sequences would also allow
a nice transition from a regular grid, $k_{xi}=\frac{2\pi}{L_{x}}\left[1,2,3,4,5,n_{x}-2,n_{x}-1\right]$
to an asymptotically logarithmic one, using the recurrence relation
on the last few nodes of the grid elements, (e.g. $k_{n_{x}}=k_{n_{x}-1}+k_{n_{x}-2}$).
Unfortunately the details of such an implementation is out of scope
of the current paper and is left to a future publication.

Also, the application of the idea to other models, such as the nested
polyhedra model, where one ``shell'' is represented by an icosahedron-dodecahedron
compound whose vertices have $x$, $y$, $z$ components which can
be written basically using combinations of $1,$$\pm\varphi$, $\pm\varphi^{2}$
so that exact triadic interactions are possible among them, can be
replaced by versions of these objects where each components can be
constructed using recurrent series, such as the Fibonacci series,
which would approach asymptotically to perfect icosahedron-dodecahedron
compounds.

\section*{ACKNOWLEDGMENT}

The authors would like to thank Dr. Benjamin Favier of IRPHE-Aix Marseille
University, for bringing this issue to our attention and the pointing
out the interest in the numerical community for such a study. The
authors would also like to thank the Isaac Newton Institute for Mathematical
Sciences, Cambridge, for support and hospitality during the programme
\textquotedblleft Anti-diffusive dynamics: from sub-cellular to astrophysical
scales\textquotedblright{} during which these discussions took place.
This work has benefited from a grant managed by the Agence Nationale
de la Recherche (ANR), as part of the program \textquoteleft Investissements
d\textquoteright Avenir\textquoteright{} under the reference (ANR-18-EURE-0014)
and has been carried out within the framework of the EUROfusion Consortium,
funded by the European Union via the Euratom Research and Training
Programme (Grant Agreement No 101052200 --- EUROfusion) and within
the framework of the French Research Federation for Fusion Studies.

%

\end{document}